# Wave vector dependence of the dynamics in supercooled metallic liquids

B. Ruta[1,2*], S. Hechler[1,3], N. Neuber[3], D. Orsi[4], L. Cristofolini[4], O. Gross[3], B. Bochtler[3], M. Frey[3], A. Kuball[3], S.S. Riegler[3], M. Stolpe[3], Z. Evenson[5], C. Gutt[6], F. Westermeier[7], R. Busch[3] and I. Gallino[3]

[1] *Univ Lyon, Université Claude Bernard Lyon 1, CNRS, Institut Lumière Matière, Villeurbanne, France*

[2] *ESRF—The European Synchrotron, CS40220, 38043 Grenoble, France*

[3] *Chair of Metallic Materials, Department of Materials Science and Engineering, Saarland University, Campus C6.3, 66123 Saarbrücken, Germany*

[4] *Dipartimento di Scienze Matematiche Fisiche ed Informatiche, Università degli Studi di Parma, Parma, Italy*

[5] *Heinz Maier-Leibnitz Zentrum (MLZ) and Physik Department, Technische Universität München, Lichtenbergstrasse 1, 85748 Garching, Germany*

[6] *Department Physik, Universität Siegen, D-57072 Siegen, Germany*

[7] *Deutsches Elektronen Synchrotron DESY, D-22607 Hamburg, Germany*

[*]Corresponding author: beatrice.ruta@univ-lyon1.fr

Abstract

We present a detailed investigation of the wave vector dependence of collective atomic motion in $Au_{49}Cu_{26.9}Si_{16.3}Ag_{5.5}Pd_{2.3}$ and $Pd_{42.5}Cu_{27}Ni_{9.5}P_{21}$ supercooled liquids close to the glass transition temperature. Using x-ray photon correlation spectroscopy in a precedent uncovered spatial range of only few interatomic distances, we show that the microscopic structural relaxation process follows in phase the structure with a marked slowing down at the main average inter-particle distance. This behavior is accompanied by dramatic changes in the shape of the intermediate scattering functions which suggest the presence of large dynamical heterogeneities at length-scales corresponding to few particle diameters. A ballistic-like mechanism of particle motion seems to govern the structural relaxation of the two systems in the highly viscous phase, likely associated to hopping of caged particles in agreement with theoretical studies.

<u>Main text</u>

The dynamics of glass-formers still attracts large interest in the scientific community as it is often considered the key to understanding the glass transition [1–5]. Considerable effort has been devoted to the comprehension of the enormous increase in viscosity, or structural relaxation time, τ, upon cooling toward the glass transition temperature, $T_g$ [5–7]. However little is still known on the microscopic mechanisms responsible for such a tremendous slowdown of viscous flow as the majority of previous studies employ macroscopic approaches. One major obstacle to overcome is the difficulty in probing the ultra-slow dynamics of glass-formers close to $T_g$ at the relevant length scale involving inter-particle interactions with both experiments and simulations.

The collective atomic motion of glass-formers can be described by the intermediate scattering function (ISF), *f(Q,t)*, which monitors the temporal evolution of the normalized density-density correlation function over a spatial scale 2π/$Q$ defined by the probed wave-vector $Q$. The long time decay of the ISF corresponds to the structural relaxation process and it can be modeled by the Kohlrausch-Williams-Watts (KWW) function $f(Q,t) = f_Q \cdot \exp[-\left(\frac{t}{\tau}\right)^{\beta}]$, with the structural relaxation time $\tau(Q)$, the shape parameter $\beta(Q)$, and the nonergodicity parameter $f_Q$ [3,8].

Studies on metallic glass-formers reports the existence of an anomalous dynamical crossover at the glass transition [9,10]. Above $T_g$, the structural relaxation process decays in a stretched exponential way with $\beta<1$. Similar decays have been attributed largely to the emergence of dynamical heterogeneities on approaching $T_g$ [3,11]. As soon as the material vitrifies, the shape of the long time decay of the ISF changes abruptly to a highly compressed form ($\beta>1$). Numerical simulations associate these compressed ISFs to an increasing number of connected icosahedral clusters on approaching $T_g$ [12].

At high temperatures, liquids undergo Brownian motion and $\tau(Q)\approx Q^{-2}$. Theoretical studies suggest a gradual weakening of this Q-dependence in supercooled liquids, due to a change in the transport mechanism from purely diffusive at high temperatures, to a combination of diffusion and hopping of caged particles on approaching $T_g$ [13]. Differently, numerical simulations show that the relaxation dynamics depends on the rigidity of the amorphous structure, being Q-independent in network glass-formers, while keeping a diffusive nature in more fragile systems [14]. This peculiar Q-independent dynamics occurs for few inter-particle distances and it has been observed also in polymeric materials and DNA-based transient networks [15–17]. Its origin appears associated to the presence of independent locally fluctuating elastic moduli on the probed length scale [15,16] and suggests the presence of a crossover to a diffusive-like dynamics at larger length scales [15,17].

A deeper understanding of the dynamics requires the experimental knowledge of the Q-dependence of the ISF in liquid and glasses. Such information is challenging to obtain. Experiments below $T_g$ are currently strongly impeded by the evolution of the dynamics due to physical aging which makes it extremely difficult to compare data acquired separately at different Qs [10,18]. The dynamics of supercooled liquids is also difficult to probe due to the relatively fast relaxation times and the weak scattering signal at the atomic scale. Due to these technical constraints, often only a single $Q$ at the maximum of the static structure factor $S(Q)$ [10,19] has been studied providing limited information on the microscopic mechanism responsible for the dynamics in supercooled liquids.

Here, we make use of recent experimental improvements in X-ray Photon Correlation Spectroscopy (XPCS) [19], and probe the Q-dependence of the relaxation dynamics at interatomic distances in two highly-viscous metallic glass-formers just a few degrees above the calorimetric $T_g$, i.e. in a previously unexplored range. The investigated systems are alloys of $Au_{49}Cu_{26.9}Si_{16.3}Ag_{5.5}Pd_{2.3}$ and $Pd_{42.5}Cu_{27}Ni_{9.5}P_{21}$ ($T_g$=574 K). When slowly cooled with of isothermal steps of 0.5 K and a cooling rate of 0.1 Kmin$^{-1}$ from the supercooled liquid, the $T_g$ of the $Au_{49}Cu_{26.9}Si_{16.3}Ag_{5.5}Pd_{2.3}$ is lowered by more than 30K [20,21], and the system exhibits a fragile-to-strong liquid-liquid transition at 389 K. In this work, we have investigated the dynamics in the low temperature strong phase ($T_g$=380 K [20]). For both studied materials, we measured the ISFs at different $Q$s around their respective maxima, $Q_m$, of the $S(Q)$ ($Q_{m,\ Au\text{-}alloy}$ = 2.80 Å$^{-1}$ and $Q_{m,\ Pd\text{-}alloy}$ = 2.87 Å$^{-1}$) and in the highly viscous state ($T_{Au\text{-}alloy}$=385.5 K, $T_{Pd\text{-}alloy}$=580 K).

The $Au_{49}Cu_{26.9}Si_{16.3}Ag_{5.5}Pd_{2.3}$ was measured at beamline ID10 at ESRF, France, and the $Pd_{42.5}Cu_{27}Ni_{9.5}P_{21}$ at beamline P10 at PETRA III, Germany (see also Supplemental Material [22]). XPCS provides information on the microscopic dynamics by monitoring the temporal evolution of the intensity fluctuations, $g_2(Q,t)$, which are related to the ISF through the Siegert relation $g_2(Q,t)=1+\gamma\cdot|f(Q,t)|^2$, with $\gamma$ the experimental contrast [19,23]. Despite the multicomponent nature of the probed alloys, in both cases, XPCS mainly measures the average dynamics coming from the noble-noble atoms correlation which dominates the scattered signal [22,24,25].

Figure 1 illustrates the two times correlation functions (TTCFs) measured in the $Au_{49}Cu_{26.9}Si_{16.3}Ag_{5.5}Pd_{2.3}$ at the low *Q* side of the maximum of the *S(Q)* (Fig. 1a), over its flank (Fig. 1 b), and at the maximum (Fig. 1c). TTCF are a time-resolved version of the standard $g_2(Q,t)$. The width of the yellow-reddish intensity along the main diagonal is proportional to $\tau(Q)$. At all *Q*s, the intensity profile remains constant with time, which is a signature of the stationary dynamics of supercooled liquids. The lower intensity at low *Q* (Fig. 1a) is a consequence of the lower scattered signal far from $Q_m$, while the small intensity fluctuations along the diagonal contour are due to the heterogeneous nature of the dynamics in supercooled liquids, which results in a distribution of microscopic distinct relaxations [11].

Fig. 2 shows normalized $g_2(Q,t)$ measured at different *Qs* together with the KWW fits $g_2(Q,t) = 1 + c\ \exp[-2\left(\frac{t}{\tau}\right)^{\beta}]$, with *c* the product between $f_Q$ and the contrast. In the probed *Q* range, *c* is found constant and the data have been normalized for clarity. Fig. 2a and 2b show normalized $g_2(Q,t)$ of the $Au_{49}Cu_{26.9}Si_{16.3}Ag_{5.5}Pd_{2.3}$ corresponding to the low and the high *Q* side of the *S(Q)*, respectively. The correlation functions visibly have distinct shapes with Q, and become more stretched at low (Fig. 2a) and high (Fig. 2b) *Qs* with respect to the shape at the maximum of the *S(Q)*. This effect is remarkable as $\beta(Q)$ decreases by almost 40% of its value, from $\beta(Q)$= 0.80±0.02 at 2.78 Å$^{-1}$ to 0.51±0.02 at 2.58 Å$^{-1}$, i.e. in a very tiny *Q* range covering only 0.2 Å$^{-1}$ (Fig. 2a). Changes in $\beta(Q)$ are visible also in the $Pd_{42.5}Cu_{27}Ni_{9.5}P_{21}$ (Fig. 2c). Here, the evolution of $\beta(Q)$ is weaker as it decreases of ≈30% of its maximum value on a *Q* range which is ≈3 times larger than the one explored for the $Au_{49}Cu_{26.9}Si_{16.3}Ag_{5.5}Pd_{2.3}$.

The structural relaxation times and shape parameters measured at all *Qs* are reported in Fig. 3 for both $Au_{49}Cu_{26.9}Si_{16.3}Ag_{5.5}Pd_{2.3}$ (left column) and $Pd_{42.5}Cu_{27}Ni_{9.5}P_{21}$ alloys (right column) and compared with the *S(Q)* measured with x-ray diffraction [25,26]. Both $\tau(Q)$ and $\beta(Q)$ mimic the *S(Q)*, showing a significant slowdown of the dynamics at $Q_m$ accompanied by a simultaneous increase in $\beta(Q)$ (panels a-b and d-e). The evolution of $\tau(Q)$ is a signature of the de Gennes narrowing usually observed in the frequency domain in high temperature liquids [27–37]. It implies that the most probable and stable interatomic configurations are those probed at $Q \approx Q_m$, which need a high degree of cooperative motion to change the atomic arrangements. Interestingly, while the *S(Q)* changes by almost a factor two in the $Au_{49}Cu_{26.9}Si_{16.3}Ag_{5.5}Pd_{2.3}$, the relative change in τ is considerably smaller. This difference is even stronger in the $Pd_{42.5}Cu_{27}Ni_{9.5}P_{21}$. Similar trends have been reported for other viscous liquids [37–39] and are likely due to the occurrence of complex mechanisms of particle motion [30]. We note that although both samples have been measured close to their $T_g$, their dynamics differ by almost an order of magnitude being of ≈$10^4$ s in the $Au_{49}Cu_{26.9}Si_{16.3}Ag_{5.5}Pd_{2.3}$. This remarkable highly viscous state is a consequence of the applied thermal protocol [20,21].

Considering the heterogeneous scenario for supercooled liquids, i.e., the existence of groups of particles relaxing with different times with respect to neighboring particles, the $g_2(Q,t)$ can be viewed as a measure of the average dynamics with $\beta(Q)$ describing the degree of such a distribution of microscopic relaxation processes [11,40]. The decrease of $\beta(Q)$ for $Q$-values smaller than that of the main peak in $S(Q)$ implies the occurrence of larger dynamical heterogeneities consistent with the notion of medium-range-order domains fluctuations. The smaller measured $Q$-values correspond generally to length scales comparable to the typical size of icosahedra clusters whose occurrence in the Pd-based alloy has been confirmed by structural studies [25,41]. The dynamics of such clusters was recently identified to strongly dominate the evolution of the ISFs in supercooled metallic liquids [12]. This behavior contrasts dramatically with that observed in metallic glasses. Below $T_g$, not only is the dynamics characterized by a compressed decay of the ISFs (i.e. $\beta>1$) and aging [10], but $\beta(Q)$ remains constant, at least in the same $Q$ range as probed here [18].

To evaluate the mechanism of particle motion we include the influence of $\beta(Q)$ on the dynamics by considering the mean relaxation time $\langle\tau(Q)\rangle = \Gamma\left(\frac{1}{\beta(Q)}\right)\frac{\tau(Q)}{\beta(Q)}$ , where Γ is the Gamma function [30]. Differently from $\tau(Q)$, $\langle\tau(Q)\rangle$ decreases with increasing $Q$ in both supercooled liquids before to evolve to a more flatten regime at larger $Q$ values (Fig. 3c and 3f). The low $Q$s regime is compatible with a $1/Q$, ballistic-like atomic motion suggesting the existence of a mechanism of hopping of caged particles in these ultra-viscous liquids [42] as in the case of some concentrated colloidal suspensions [43], gels [44,45] and in numerical simulations of ortho-terphenyl [46]. Interestingly, the crossover between the two regimes occurs at different $Q$s in the two alloys. In the $Au_{49}Cu_{26.9}Si_{16.3}Ag_{5.5}Pd_{2.3}$, the ballistic-like regime persists up to $Q \approx Q_m$, while in the $Pd_{42.5}Cu_{27}Ni_{9.5}P_{21}$ it ends in correspondence of the onset of the main structural peak in the $S(Q)$. Such a difference is likely due to the huge reduction in $\beta(Q)$ at low $Q$s for the $Au_{49}Cu_{26.9}Si_{16.3}Ag_{5.5}Pd_{2.3}$, of about 40% of its value at $Q_m$ (Fig. 3a), which dominates the $\langle\tau(Q)\rangle$ evolution. In the very same $Q$ range, $\beta(Q)$ of $Pd_{42.5}Cu_{27}Ni_{9.5}P_{21}$ decreases of only 15% of its value at $Q_m$ (Fig. 3d). In this range, $\langle\tau(Q)\rangle$ is mainly controlled by the $Q$-dependence of the relaxation time and decreases with lower $Q$, following in phase the $S(Q)$. The distinct crossovers between the two dynamical regimes could be associated to the different structural detailed of the two alloys and in particular to the stiffer nature of the Pd-based alloy [25] with respect to the microscopically softer structure of the $Au_{49}Cu_{26.9}Si_{16.3}Ag_{5.5}Pd_{2.3}$ [26].

We note that in the $Au_{49}Cu_{26.9}Si_{16.3}Ag_{5.5}Pd_{2.3}$ alloy the evolution of $\langle\tau(Q)\rangle$ comes from the marked $Q$-dependence of both $\beta(Q)$ and $\tau(Q)$ in the probed $Q$ range, confirmed by the distinct corresponding ISFs (Fig. 2a). In stark contrast, the sub-quadratic dependence of $\langle\tau(Q)\rangle$ in the $Pd_{42.5}Cu_{27}Ni_{9.5}P_{21}$ occurs in a $Q$ range where the corresponding KWW parameters are almost constant with $Q$s (Fig. 3d and 3e). The slightly decreases of $\beta(Q)$ at low $Q$s is then likely to be the reason of the still marked $Q$-dependence of the mean relaxation time in this range. Whether such dependence is real or not is however difficult to establish. As shown in Fig. 4, the decay of the corresponding ISFs almost overlap with $Q$, being the relaxation time basically constant at low $Q$s, i.e. far from the structural contribution (inset of Fig. 4). Furthermore, the scattering of the data at short times introduces some noise in the evaluation of $\beta(Q)$ and $\tau(Q)$ which could affect the probed $Q$-dependencies. This lower signal to noise ratio (SNR) in the $Pd_{42.5}Cu_{27}Ni_{9.5}P_{21}$ data at small $Q$s is due to the limited capabilities of the XPCS technique at synchrotrons of 3$^{rd}$ generation and the relatively fast dynamics of this liquid which is almost a factor 5 faster than in the Au-based alloy (see also Supplemental Material [22]). Hence, measurements with improved statistics or in a broader $Q$ range will be necessary to confirm the observed sub-quadratic collective dynamics or instead point to a $Q$-independent relaxation mode between the hydrodynamics limit and the inter-particle distance [14], being the hydrodynamic limit around few μm for the Pd-alloy [47].

In conclusion, we reported a detailed investigation of the atomic motion in highly viscous alloys close to $T_g$. Both $\tau(Q)$ and $\beta(Q)$ follow in phase the $S(Q)$. While the slowdown of the dynamics is due to the well-known de Gennes narrowing [29], the substantial decrease of $\beta(Q)$ at small $Q$s suggests the occurrence of large dynamical heterogeneities possibly related to medium-range-order domains fluctuations. A similar but smaller $\beta(Q)$ evolution is predicted by the mode coupling theory for hard spheres [48,49], and has been reported also in high temperature polymeric liquids [30] and in simulations of supercooled water [50]. In our case, the collective motion of both liquids follows a subquadratic dependence from the probed wavevector, suggesting the occurrence of a complex mechanism of diffusion and hopping responsible for the particle motion [42].

The dynamics of metallic glass-formers across $T_g$ has been often compared to that of nanoparticles probes in supercooled molecular and polymeric liquids [51,52]. In both cases, the glass transition is accompanied by the emerging of compressed ISFs and distinct aging regimes. This analogy does not extend to the $Q$-dependence of the dynamics. While the dynamics of nanoparticles probes evolve from diffusive with stretched ISFs to ballistic-like with compressed ISFs on approaching the glassy state, our results show the existence of a distinct mechanism of particle motion, which appears related to the details of the amorphous structure in the viscous phase. This different dynamics contrasts also with the diffusive dynamics observed in Lennard Jones systems which are often considered similar to metallic alloys [14]. Studies of the dynamics across $T_g$ and in a broader $Q$ range will definitely help on elucidating the nature of the glass transition. This will be soon possible thanks to the current upgrade of coherent X-ray sources [53,54].

The authors thank W. Kob, L. Rovigatti and F. Yang, for helpful discussions. We gratefully thank ESRF and DESY for providing beamtime. Y. Chushkin and M. Sprung are acknowledged for providing the codes for the data analysis. H. Vitoux, K. L'Hoste and F. Zontone are acknowledged for the support during the XPCS measurements at ESRF. I. Gallino acknowledges the DFG for financial support from Grant No. GA 1721/2-2 and B. Ruta acknowledges the CNRS for the PICS 278566 funding. Ralf Busch acknowledges financial support from the German Federation of Industrial Research Associations (AiF/IGF) through Project No.17716N and Heraeus Holding GmbH for the gold, palladium and silver supply.

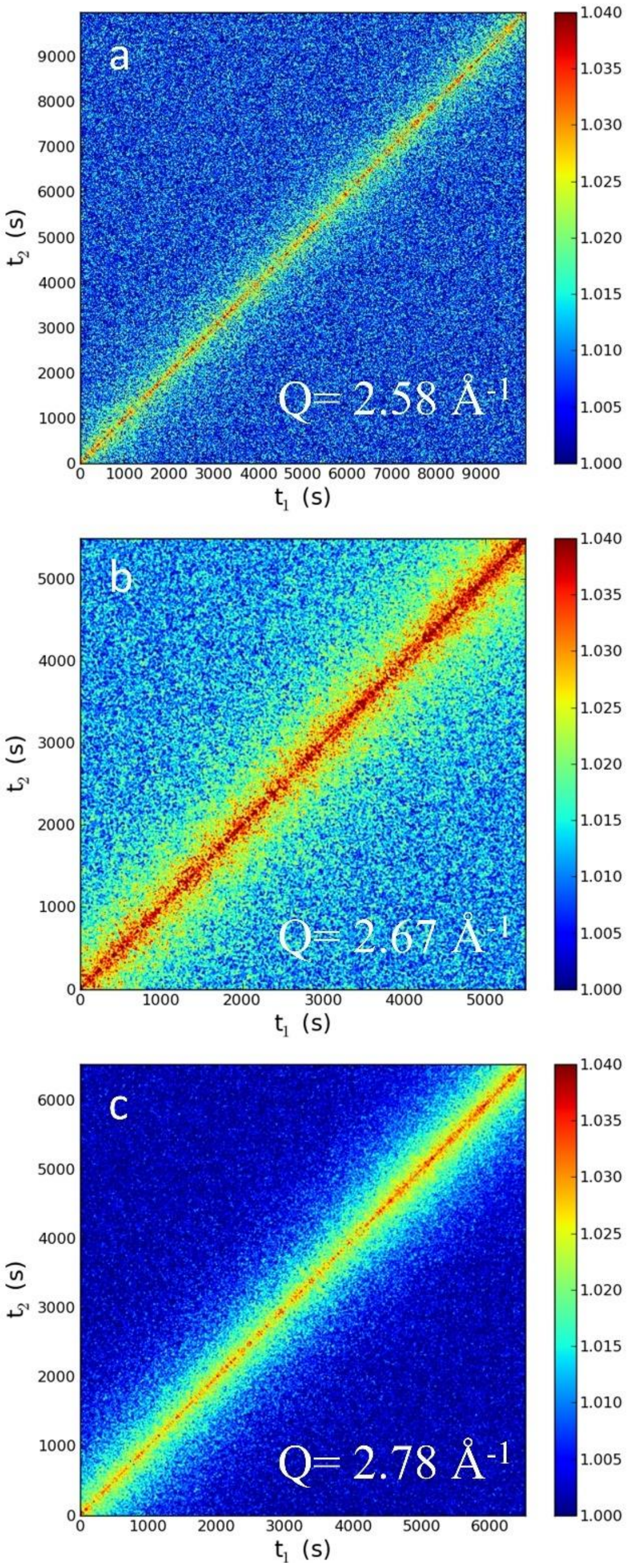

**Figure 1:** TTCF at 385.5 K in $Au_{49}Cu_{26.9}Si_{16.3}Ag_{5.5}Pd_{2.3}$ for three different *Q*s (2.58 Å$^{-1}$ (**a**), 2.67 Å$^{-1}$ (**b**), 2.78 Å$^{-1}$ (**c**)).

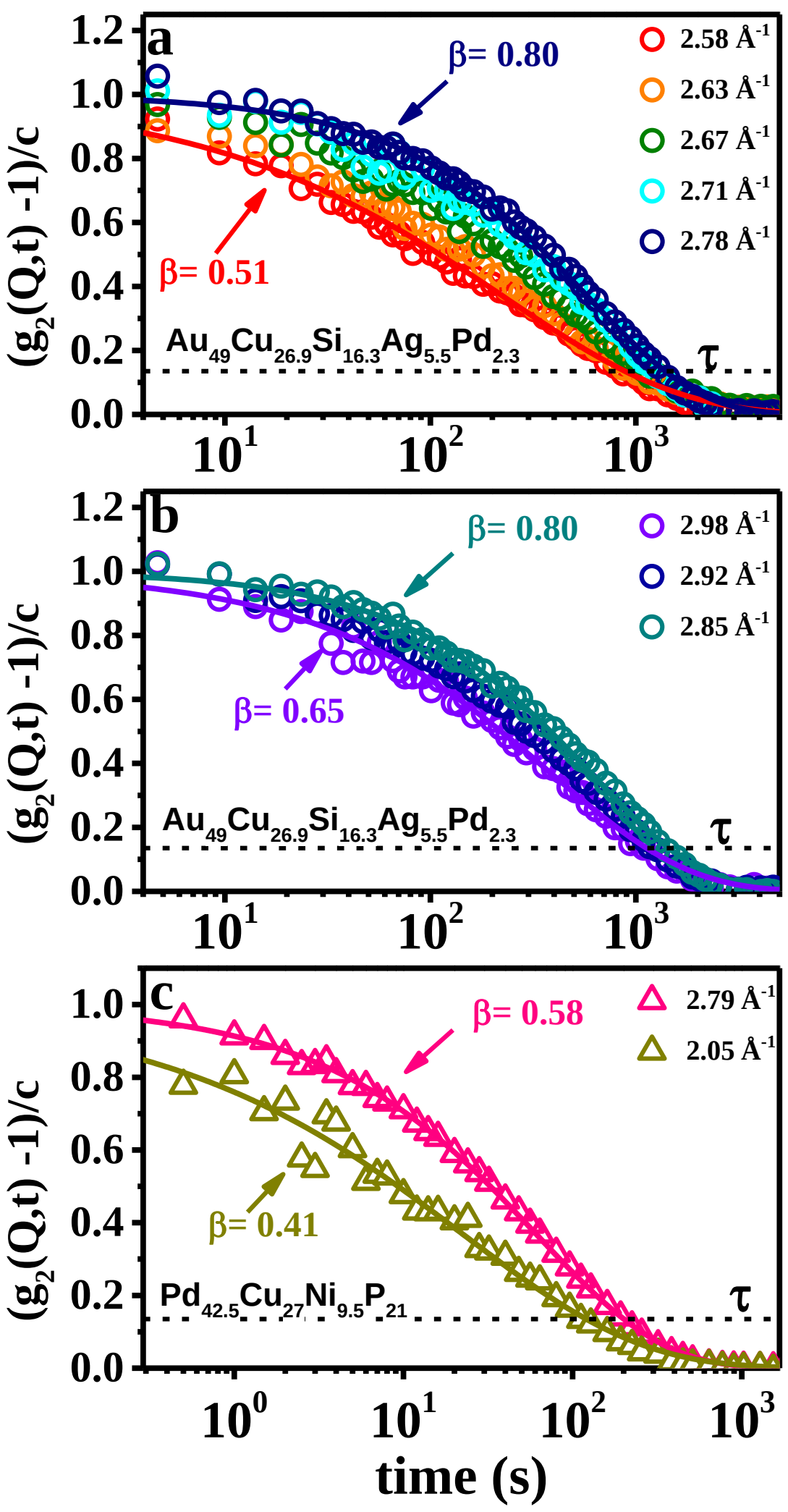

**Figure 2:** Normalized intensity autocorrelation functions for the low (**a**) and the high (**b**) $Q$ side of the main maximum of the $S(Q)$ in $Au_{49}Cu_{26.9}Si_{16.3}Ag_{5.5}Pd_{2.3}$ at T=385.5 K, and (**c**) at the smallest probed $Q$ and at the $Q_{m,Pd\text{-}alloy}$ for $Pd_{42.5}Cu_{27}Ni_{9.5}P_{21}$ at T=580 K.

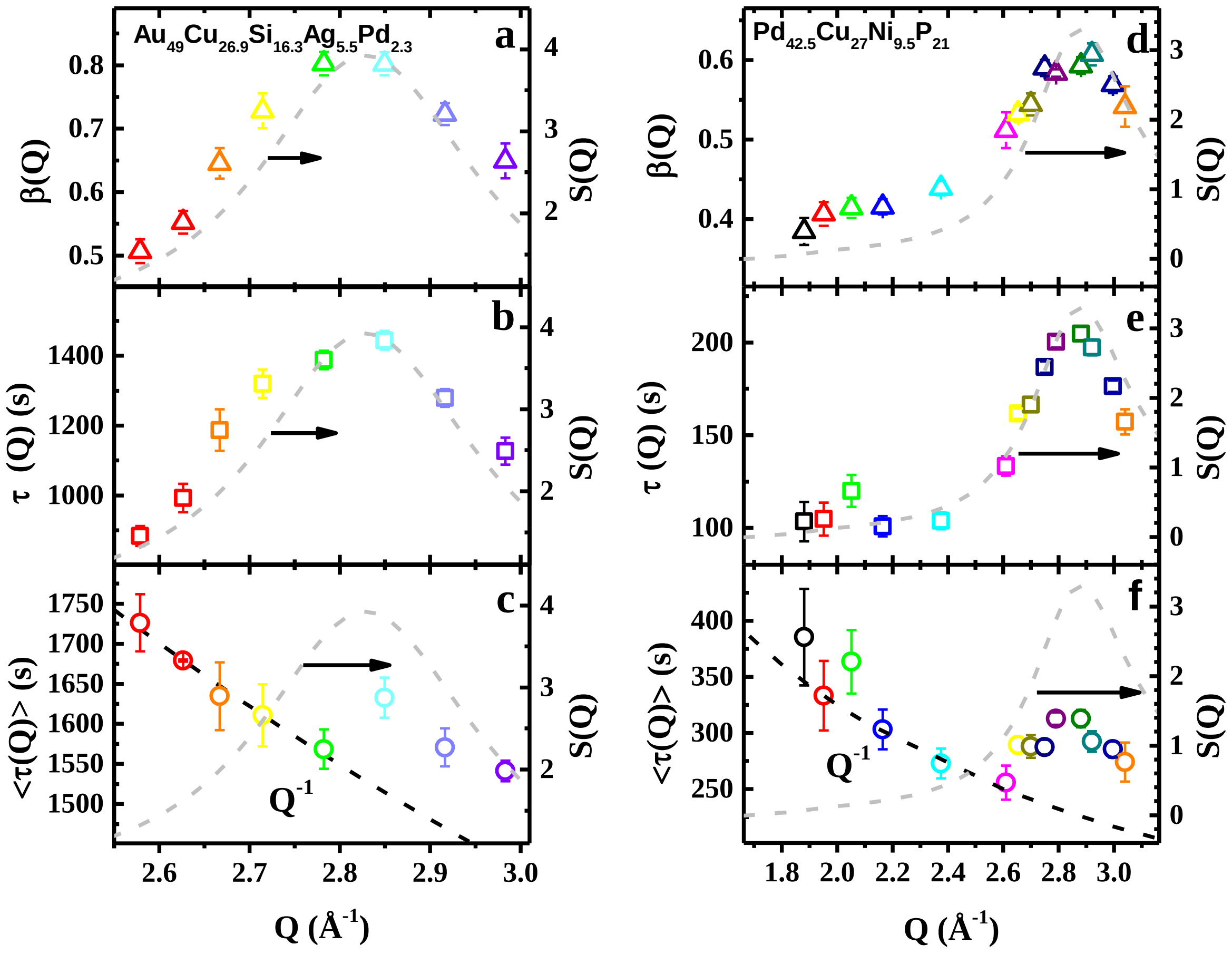

**Figure 3:** *τ(Q)* (**a, d**), *β(Q)* (**b, e**) and average relaxation time *<τ(Q)>* (**c, f**) for $Au_{49}Cu_{26.9}Si_{16.3}Ag_{5.5}Pd_{2.3}$ (left column) and $Pd_{42.5}Cu_{27}Ni_{9.5}P_{21}$ (right column). The colors of the Au-based alloy correspond to the curves in Fig. 2. The dashed black lines in (**c** and **f**) show a ballistic-like dynamics. In all panels, the dashed grey line is the corresponding *S(Q)*.

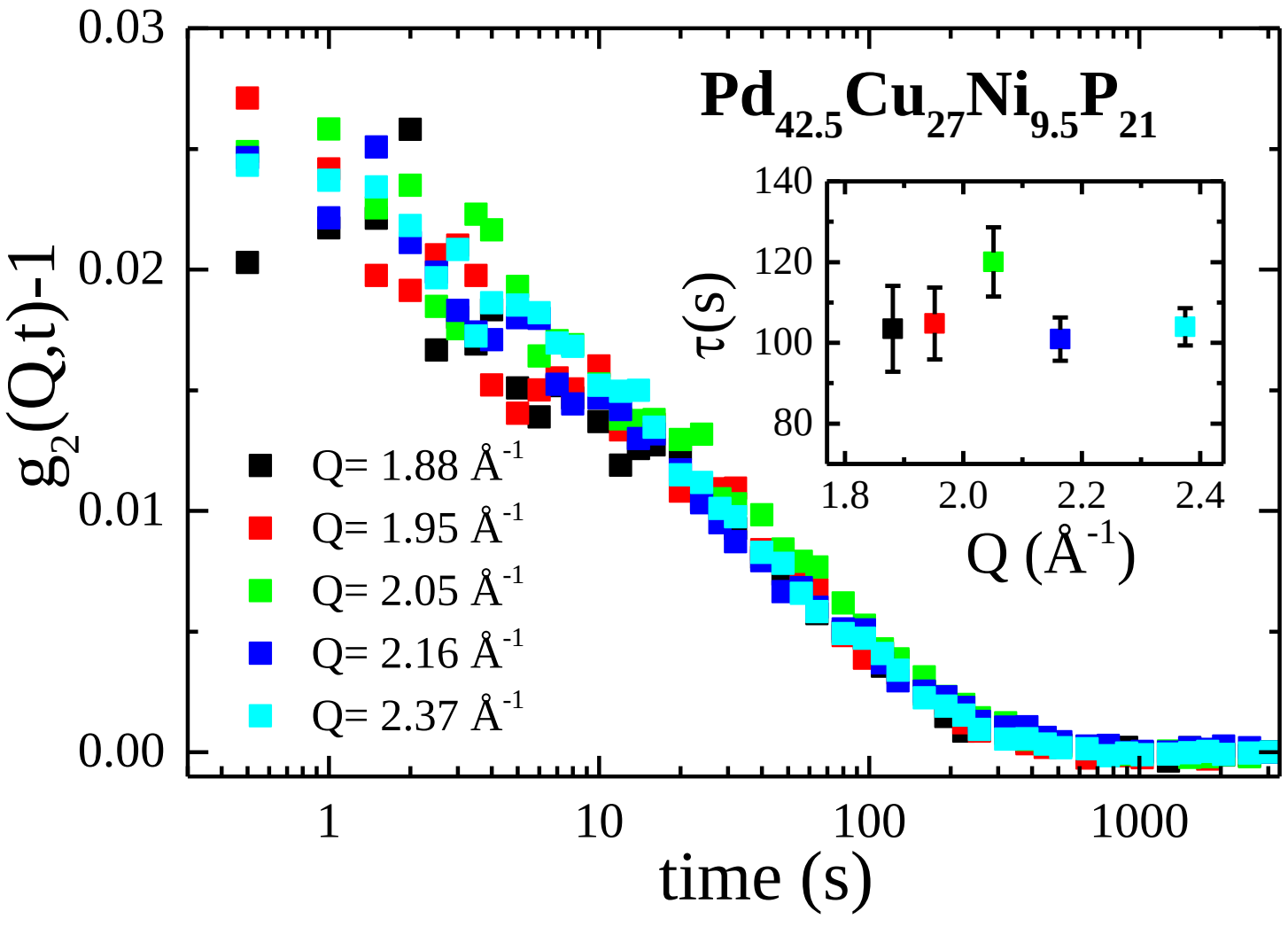

**Figure 4:** Intensity autocorrelation functions measured at the small *Qs* in the $Pd_{42.5}Cu_{27}Ni_{9.5}P_{21}$ at T=580 K. The inset shows the corresponding τ(Q).